\documentclass[twocolumn,english]{revtex4}
\usepackage[T1]{fontenc}
\usepackage[latin1]{inputenc}
\usepackage{amsmath}
\usepackage{graphicx}
\usepackage{amssymb}

\makeatletter



\usepackage{babel}
\makeatother
\begin{document}

\title{Non adiabatic quantum search algorithms}

\author{A. Pérez$^{1}$ and A. Romanelli$^{1,2}$}

\affiliation{$^{1}$Departament de Física Te\`{o}rica and IFIC, Universitat de
Val\`{e}ncia-CSIC \\
Dr. Moliner 50, 46100-Burjassot, Spain\\
 $^{2}$ Instituto de Física, Facultad de Ingeniería, Universidad
de la Rep\'{u}blica, \\
C.C. 30, C.P. 11000, Montevideo, Uruguay}

\begin{abstract}
We present two new continuous time quantum search algorithms similar
to the adiabatic search algorithm, but now without an adiabatic evolution.
We find that both algorithms work for a wide range of values of the
parameters of the Hamiltonian, and one of them has, as an additional
feature that, for values of time larger than a characteristic one,
it will converge to a state which can be close to the searched state. 
\end{abstract}
\maketitle

\section{Introduction}

Quantum computation has attracted the attention of researchers from
several different areas \cite{Chuang}. This field of knowledge presents
new scientific challenges to learn how to work with quantum properties
to obtain more efficient algorithms. However, relatively few quantum
algorithms were created; among them, Shor's and Grover's \cite{Shor,Grover}
algorithms are the best known. Grover's search algorithm locates a
marked item in an unsorted list of $N$ elements in a number of steps
proportional to $\sqrt{N}$, instead of $O(N)$ as in the classical
case. It performs a unitary transformation of the initial quantum
state so as to increase the likelihood that the marked state of interest
will be measured at the output (amplification technique). It has been
proved that there are neither quantum nor classical algorithms that
can perform faster such an unstructured search \cite{Boyer}. This
search algorithm has also a continuous time version \cite{Farhi}
that has been described as the analogue of the original Grover algorithm.
From this continuous time version, and using the quantum adiabatic
theorem, adiabatic search algorithms have been developed \cite{Farhi2,Farhi3,Roland},
that consist in guessing a time-dependent Hamiltonian whose dynamics
evolves slowly enough so that it remains always near its instantaneous
ground state. They solve the search problem in a time proportional
to $\sqrt{N}/\delta$, where $\delta$ is a precision parameter that
depends on the energy difference between the two lowest states.

Another way to generate a continuous time quantum search algorithm
\cite{alejo} has been recently developed, that finds a discrete eigenstate
of a given Hamiltonian $H_{0}$. This algorithm behaves like Grovers's,
and explicitly shows that the search algorithm is essentially a resonance
phenomenon between the initial and the searched states \cite{Grover2}.

In this work we present two new continuous time search algorithms
that are controlled by a time dependent Hamiltonian, similarly to
the case of the quantum adiabatic search algorithm, but now the evolution
is non adiabatic; then it is not necessary to impose slowness to the
dynamics in order to preserve the system in the fundamental state.
These algorithms provide new insights to search algorithms, in particular:
a connection between the resonant and the adiabatic search algorithms
(for the first case), or the possibility to generate a new type of
search algorithm in which one does not need to pick up a particular
instant of time when the measure has to be performed (for the second
case), provided the parameters characterizing the Hamiltonian are
conveniently chosen. In this second case, one reaches an asymptotic
form for the searching probability, which is rather independent on
the size of the database $N$, at the cost of increasing the energy
resources in the Hamiltonian as $N$ grows.

The paper is organized as follows. In the next two sections we develop
the two models of non adiabatic algorithms. In the last section we
draw the conclusions of this work.

\section{Non adiabatic algorithm I}

Consider $N$ items in the database, each associated with a vector
in the complete orthonormal set $\left\{ |n\rangle,\text{ }n=1,2,...,N\right\} $
in a Hilbert space. Let us call $|s\rangle$ the unknown searched
state that is associated with the marked item belonging to the previous
group. We assume that the initial state is the symmetric normalized
state \begin{equation}
\left\vert \psi_{0}\right\rangle =\frac{1}{\sqrt{N}}\sum\limits _{n=1}^{N}|n\rangle,\label{Initialstate}\end{equation}
 and define the two Hamiltonians\begin{equation}
H_{0}=I-\left\vert \psi_{0}\right\rangle \left\langle \psi_{0}\right\vert ,\label{h0}\end{equation}
 \begin{equation}
H_{s}=I-\left\vert s\right\rangle \left\langle s\right\vert ,\label{h1}\end{equation}
 where $I$ is the identity matrix, their ground states being $\left\vert \psi_{0}\right\rangle $
and $\left\vert s\right\rangle $ respectively. The algorithm is built
on the following time-dependent Hamiltonian \begin{equation}
H(t)=f(t)H_{0}+g(t)H_{s}\,,\label{hamiltonian}\end{equation}
 where $f(t)$ and $g(t)$ are time dependent functions that will
be defined later. Notice that\ $H_{s}$ plays the equivalent role
to 'marking' the searched state in Grover's algorithm. The goal of
the search algorithm is to change $\left\vert \psi_{0}\right\rangle $
into $|s\rangle$ or some approximation there of, following the dynamics
generated by the Schrödinger equation. In this problem, we can restrict
the analysis to the two-dimensional space spanned by $|s\rangle$
and $|p\rangle=$ $\frac{1}{\sqrt{N-1}}\sum\limits _{n=1,n\neq s}^{N}|n\rangle$.
The wave function is then expressed as\begin{equation}
\left\vert \psi(t)\right\rangle =a_{s}(t)|s\rangle+a_{p}(t)|p\rangle\,,\label{funcion1}\end{equation}
 for some $a_{s}(t)$, $a_{p}(t)$ such that \ $\left\vert a_{s}(t)\right\vert ^{2}+\left\vert a_{p}(t)\right\vert ^{2}=1$
with $a_{s}(0)=\sqrt{1/N}$ and $a_{p}(0)=\sqrt{1-1/N}$. In the $\{|s\rangle$,
$|p\rangle\}$ basis we have the following matrix for the Hamiltonian\ \ \begin{equation}
H(t)=\frac{1}{N}\left[\begin{array}{cc}
\left(N-1\right)f(t) & -\sqrt{N-1}f(t)\\
-\sqrt{N-1}f(t) & \text{ \ \ \ }f(t)+Ng(t)\end{array}\right].\label{matriz1}\end{equation}
 The above matrix can be rewritten under the form

\begin{equation}
H(t)=\frac{1}{2}(f+g)I+\frac{1}{2}\omega(t)\vec{n}\cdot\vec{\sigma}\label{matriz2}\end{equation}
 where%
\footnote{For brevity, we omit in some equations the dependence of the functions
$f$, $g$ and $\theta$ on $t$.%
} $\omega(t)=\sqrt{(f-g)^{2}+4fg/N}$, $\vec{n}=\frac{1}{\omega(t)}(-2\sqrt{N-1}f/N,0,f-g-2f/N)$
is an unitary vector and $\vec{\sigma}$ stands for the Pauli matrices.
Defining $\vec{n}=(\sin\theta,0,\cos\theta)$, one can easily obtain
the functions $f$ and $g$ as a function of $\theta$, with the following
result:

\begin{eqnarray}
f(t) & = & -\frac{N}{2\sqrt{N-1}}\omega(t)\sin\theta\label{ftheta}\\
g(t) & = & -\frac{N}{2\sqrt{N-1}}\omega(t)\cos(\theta+\beta)\label{gtheta}\end{eqnarray}
 with $\sin\beta\equiv(2-N)/N$, $\cos\beta\equiv2\sqrt{N-1}/N$.

The first term in Eq. (\ref{matriz2}) is proportional to the identity,
and therefore amounts to a common (time dependent) phase that can
be ignored if one only wants to evaluate probabilities. We will concentrate
on the second term, which has eigenvalues $E_{\pm}(t)=\pm\frac{1}{2}\omega(t)$
with corresponding time-dependent eigenvectors $|E_{+},t>=(\cos\frac{\theta}{2},\sin\frac{\theta}{2})$
and $|E_{-},t>=(-\sin\frac{\theta}{2},\cos\frac{\theta}{2})$, with
respect to the basis $\{|s\rangle,p\rangle\}$. In this form, it becomes
evident that the evolution originated from the Hamiltonian amounts
to a (time-dependent) rotation in the space spanned by the states
$\{|s\rangle,p\rangle\}$ with the goal of maximizing the probability
of the $|s\rangle$ state.

The Schrödinger equation in this basis, in units such that $\hbar=1$,
becomes\begin{equation}
\left[\begin{array}{c}
\frac{da_{s}(t)}{dt}\\
\frac{da_{p}(t)}{dt}\end{array}\right]=-iH(t)\left[\begin{array}{c}
a_{s}(t)\\
a_{p}(t)\end{array}\right].\label{scho1}\end{equation}
 Now we take the following steps: first, we change to a new basis
where this Hamiltonian is diagonal; second, we solve the Schrödinger
equation in that basis, and third, we return to the original basis,
where we are searching the state $|s\rangle$. The wave function,
$\left\vert \psi(t)\right\rangle $ can be expressed as a combination
of the time-dependent eigenstates $|E_{+},t\rangle$, $|E_{-},t\rangle,$

\begin{eqnarray}
\left\vert \psi(t)\right\rangle  & = & a_{+}(t)\exp(-i\int\limits _{0}^{t}E_{+}(t)dt)|E_{+},t\rangle\nonumber \\
 &  & +a_{-}(t)\exp(-i\int\limits _{0}^{t}E_{-}(t)dt)|E_{-},t\rangle.\label{funcion2}\end{eqnarray}

We have two expressions for the wave function, one in the $\{|s\rangle$,
$|p\rangle\}$ basis, Eq. (\ref{funcion1}), and the other one in
the $\{|E_{+},t\rangle,|E_{-},t\rangle\}$ basis, Eq. (\ref{funcion2}).
The relation between both basis can be expressed as a relation between
its coefficients, that is\begin{equation}
\left[\begin{array}{c}
a_{s}(t)\\
a_{p}(t)\end{array}\right]=U^{\dagger}(t)\left[\begin{array}{c}
a_{+}(t)\exp(-i\int\limits _{0}^{t}E_{+}(t)dt)\\
a_{-}(t)\exp(-i\int\limits _{0}^{t}E_{-}(t)dt)\end{array}\right]\text{,}\label{aes1}\end{equation}
 where \begin{equation}
U^{\dagger}(t)=\left(\begin{array}{cc}
\cos\frac{\theta}{2} & -\sin\frac{\theta}{2}\\
\sin\frac{\theta}{2} & \cos\frac{\theta}{2}\end{array}\right),\label{cambiobase}\end{equation}
 The Schrödinger equation in the new coordinates is\begin{equation}
\left[\begin{array}{c}
\frac{da_{+}(t)}{dt}\\
\frac{da_{-}(t)}{dt}\end{array}\right]=-M^{\dagger}(t)U(t)\frac{dU^{\dagger}(t)}{dt}M(t)\left[\begin{array}{c}
a_{+}(t)\\
a_{-}(t)\end{array}\right],\label{schro}\end{equation}
 where $M_{11}(t)=\exp(-i\int\limits _{0}^{t}E_{+}(t)dt)$, $M_{22}(t)=\exp(-i\int\limits _{0}^{t}E_{+}(t)dt)$
and $M_{12}(t)=M_{21}(t)=0$. From here, one easily arrives to\begin{align}
\frac{da_{+}(t)}{dt} & =-\Omega(t)\text{ \ }a_{-}(t),\label{eqdiferencial1}\\
\frac{da_{-}(t)}{dt} & =\Omega^{\ast}(t)\text{ }a_{+}(t),\label{eqdiferencial2}\end{align}
 where \begin{equation}
\Omega(t)\equiv-\frac{1}{2}\dot{\theta}\exp(i\int\limits _{0}^{t}\omega(t)dt),\label{omega}\end{equation}
 with $\dot{\theta}=\frac{d\theta}{dt}$. Alternatively, we can rewrite
Eq. (\ref{omega}) as

\begin{eqnarray}
\Omega(t) & = & \frac{\sqrt{N-1}}{N\omega²}\left(\frac{dg(t)}{dt}f(t)-g(t)\frac{df(t)}{dt}\right)\nonumber \\
 & \times & \exp(i\int\limits _{0}^{t}\omega(t)dt);\label{defomega}\end{eqnarray}
 we shall take $f(0)=1$ and $g(0)=0$ in this section, then the initial
condition in the new coordinates are $a_{+}(0)=0$ and $a_{-}(0)=1$.
Up to this point, our treatment of the problem is similar to that
of the adiabatic algorithm. Now to proceed further we shall choose
the function $\theta(t)$ (or, equivalently, the functions $f(t)$
and $g(t)$) for the non adiabatic approach that has similarities
with the resonant search algorithm. As seen in Eq. (\ref{omega}),
if we choose $\theta(t)$ so as to cancel the time dependence of the
modulus of $\Omega$, the system Eqs. (\ref{eqdiferencial1}, \ref{eqdiferencial2})
will have an oscillatory solution between the amplitudes $a_{+}(t)$
and $a_{-}(t)$, with a period proportional to $\sqrt{N}$, which
can be identified with the Grover search time. This situation reminds
us of the resonant search algorithm \cite{alejo}, but now these amplitudes
are not the amplitudes $a_{s}(t)$ and $a_{p}(t)$.

In order to solve analytically the system Eqs. (\ref{eqdiferencial1},
\ref{eqdiferencial2}) we shall impose the conditions:

\begin{equation}
\frac{dg(t)}{dt}f(t)-g(t)\frac{df(t)}{dt}\equiv\varepsilon\left(\alpha t+\gamma\right)^{2},\label{rela1}\end{equation}

\begin{equation}
\omega(t)=\left\vert \alpha t+\gamma\right\vert .\label{rela2}\end{equation}
 In these expressions $\varepsilon$ is\ a coupling parameter between
the states $a_{+}(t)$ and $a_{-}(t)$, $\alpha$ is the velocity
parameter associated to the energy gap and $\gamma=1$ is dictated
by our choice of $f(0)$ and $g(0)$. One can check that Eqs. (\ref{rela1},\ref{rela2})
are equivalent to imposing

\begin{equation}
\dot{\theta}=2\Omega_{0}\label{deftheta}\end{equation}
 with $\Omega_{0}$ a constant, which we rewrite as $\Omega_{0}=\frac{\sqrt{N-1}}{N}\varepsilon$.
In this way conditions (\ref{rela1},\ref{rela2}) simply imply both
a \textit{mixing angle} $\theta$(t) and a \textit{gap} $\omega(t)$
which evolve linearly with time, and thus determine the time evolution
of functions $f(t)$ and $g(t)$, see Eqs. (\ref{ftheta},\ref{gtheta}).

We can decouple Eqs. (\ref{eqdiferencial1}, \ref{eqdiferencial2})
to obtain a differential equations for $a_{{-}}(t)$ (and similarly
for $a_{{+}}(t)$) \begin{equation}
\frac{d^{2}a_{-}}{dt^{2}}+i\omega(t)\frac{da_{-}}{dt}+\Omega_{0}{}^{2}a_{-}=0.\label{couple}\end{equation}

This equation can be solved in the same way as was done in \cite{Zener}.
The change of variable\begin{equation}
W=\exp\left[\frac{i}{4}(\alpha t^{2}+2\gamma t)\right]a_{-},\label{cdv}\end{equation}
 leads to \begin{equation}
\frac{d^{2}W}{dz^{2}}+\left(\eta+\frac{1}{2}-\frac{1}{4}z^{2}\right)W=0,\label{Whittaker}\end{equation}
 where\begin{equation}
z=\sqrt{\alpha}\exp(-i\pi/4)(t+\gamma/\alpha)\label{zeta}\end{equation}
 and\begin{equation}
\eta=i\frac{\Omega_{0}^{2}}{\alpha}.\label{eta}\end{equation}
 The solutions of the Eqs. (\ref{Whittaker}) are the parabolic cylinder
function $D_{\eta}(z)$ \cite{Gradshteyn}. In our case the general
solution is\begin{equation}
W(z)=A_{1}D_{\eta}(z)+A_{2}D_{\eta}(-z),\label{whitt2}\end{equation}
 where the coefficients are determined by the initial conditions $a_{-}(0)=1$,
and $\frac{da_{-}(0)}{dt}=0$. These coefficients are\begin{equation}
A_{1}=\frac{D_{\eta-1}(-z_{0})}{D_{\eta-1}(z_{0})D_{\eta}(-z_{0})+D_{\eta}(z_{0})D_{\eta-1}(-z_{0})},\label{n3}\end{equation}
 \begin{equation}
A_{2}=\frac{D_{\eta-1}(z_{0})}{D_{\eta-1}(z_{0})D_{\eta}(-z_{0})+D_{\eta}(z_{0})D_{\eta-1}(-z_{0})}.\label{n4}\end{equation}
 Finally, the amplitude $a_{+}(t)$ can be calculated using the above
result for $a_{-}(t)$ and Eq. (\ref{eqdiferencial2}).

Let us discuss in more detail the qualitative behavior of the results
we have obtained so far. For large $N$ and finite $\varepsilon$
in such a way that $\Omega_{0}\simeq\frac{\varepsilon}{\sqrt{N}}\ll1$
it follows from the Eqs. (\ref{cdv}, \ref{zeta}, \ref{eta}, \ref{whitt2},
\ref{n3}, \ref{n4}) that $\left\vert a_{{-}}(t)\right\vert \simeq1$,
$\left\vert a_{{+}}(t)\right\vert \simeq0$. On the other hand, for
large $N$ one can approximate $\theta\simeq2\Omega_{0}t$. Then using
Eq. (\ref{aes1}), the following approximation for the probabilities
of the searched and the orthogonal states are obtained \begin{equation}
P_{s}(t)\simeq\sin^{2}(\Omega_{0}t),\label{ps}\end{equation}
 \begin{equation}
P_{p}(t)\simeq\cos^{2}(\Omega_{0}t),\label{pp}\end{equation}
 which are valid whenever $t$ satisfies $(\alpha t+\gamma)>0$. Note
that the Eqs. (\ref{ps}, \ref{pp}) are independent of the value
of $\alpha$ if the previous conditions are verified. Then, if we
let the system evolve during a time $\tau\equiv\frac{\pi\sqrt{N}}{2\varepsilon}$,
and we measure immediately after that, the probability to obtain the
searched state is equal to one. In this case our method behaves qualitatively
like Grover's. The parameter $\varepsilon$ allows us for a faster
search (relatively to the standard Grover's algorithms): one can even
obtain a characteristic time $\tau\sim1$. This speedup is allowed
because the energy scale in the Hamiltonian, defined by the functions
$f(t)$ and $g(t)$ is large enough (c.f. Eq. (22) in \cite{Das}),
provided that $\Omega_{0}\ll1$. For concreteness, we will adopt the
value $\varepsilon=1$. Additionally we have recovered our interpretation
of the search algorithm as a quantum resonance between states \cite{alejo,alejo1,alejo2,alejo3};
now the resonance is between the searched and the orthogonal states.

\begin{figure}[h]
\begin{centering}
\includegraphics[scale=0.35]{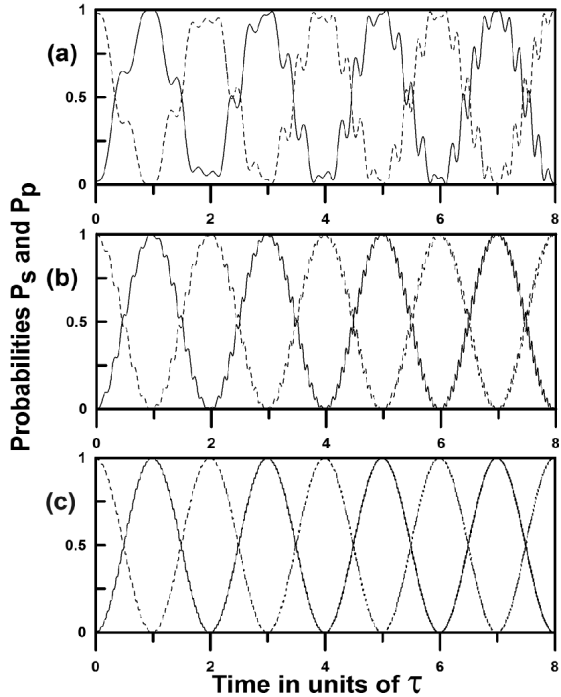} 
\par\end{centering}

\caption{Probabilities as a function of time in units of the characteristics
time $\tau\equiv\frac{\pi\sqrt{N}}{2\varepsilon}$. The the full line
corresponds to the searched state $P_{s}$, the dashed line to the
orthogonal state $P_{p}$. In all cases $\alpha>0$, but the probabilities
are independent of the value of $\alpha$. The sizes of the searched
set are: (a) $N=50$, (b) $N=500$, (c) $N=5000.$}

\label{ps1} 
\end{figure}

The above result shows that the non adiabatic algorithm works correctly
for $\alpha>0$ (remember that $\gamma=1$). We have verified this
situation for several values of $N$ and $\alpha>0$ using the exact
equation in Fig. \ref{ps1}. The figure shows a periodic behavior
with the Grover characteristic time and the correctness of the approximation
in Eqs. (\ref{ps}, \ref{pp}) as $N$ increases. %

\begin{figure}[h]

\begin{centering}
\includegraphics[scale=0.35]{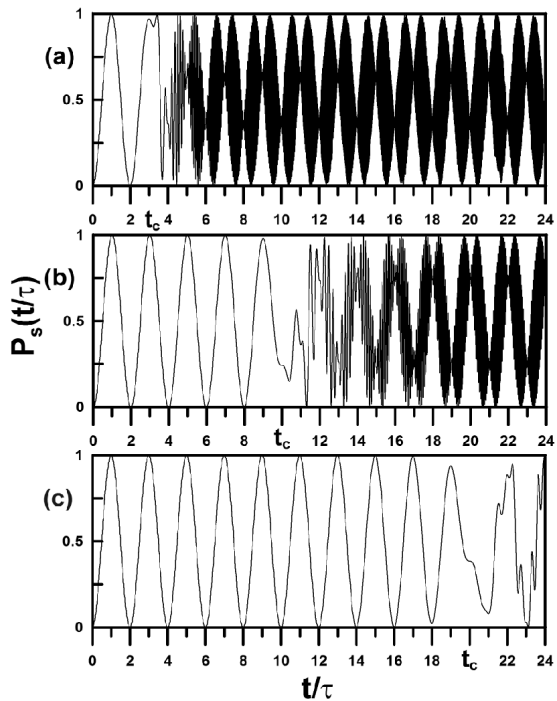} 
\par\end{centering}

\caption{Probability of the searched state $P_{s}$ as a function of time.
$N=5000$, in (a) $\alpha=-0.31\gamma/\tau$, in (b) $\alpha=-0.10\gamma/\tau$
in (c) $\alpha=-0.05\gamma/\tau$. Note that the values $t_{c}$ change
inversely with $\alpha$.}

\label{ps2} 
\end{figure}

When $\alpha=0$ (see Eq. (\ref{couple})) $a_{-}$ and $a_{+}$ can
be easily obtained analytically

\begin{eqnarray}
a_{-}(t) & = & \exp(-i\frac{\gamma}{2}t)\biggl[\cos\phi(t)\nonumber \\
 & + & \frac{i\gamma}{\sqrt{\left(\frac{\gamma}{2}\right)^{2}+\Omega_{0}^{2}}}\sin\phi(t)\biggr]\label{am1}\end{eqnarray}

\begin{equation}
a_{+}(t)=\frac{\Omega_{0}\exp(-i\frac{3\gamma}{2}t)}{\sqrt{\left(\frac{\gamma}{2}\right)^{2}+\Omega_{0}^{2}}}\sin\phi(t)\label{am2}\end{equation}
 with $\phi(t)=\sqrt{\left(\frac{\gamma}{2}\right)^{2}+\Omega_{0}^{2}}t$.
From these expressions, and for large $N$, $\Omega_{0}\ll1$, $\left\vert a_{+}(t)\right\vert \simeq0$,
$\left\vert a_{-}(t)\right\vert \simeq1$ and using the same arguments
as before, it can be shown that the search algorithm also works in
this case. For $\alpha<0$ the behavior of the system is quite more
complex. If, additionally, $\left\vert \alpha\right\vert \simeq0$
(then both $\left\vert z_{0}\right\vert $ and $\left\vert z\right\vert $
go to infinity) using Eqs. (\ref{zeta}, \ref{whitt2}, \ref{n3},
\ref{n4}, \ref{aes1}) and the asymptotic property of the parabolic
cylinder functions, it can be shown that $P_{-}(t)=\left\vert a_{-}(t)\right\vert ^{2}\simeq1$
as before, then for large $N$ and $\Omega_{0}<1$ Eqs. (\ref{ps},
\ref{pp}) are again obtained, and the search algorithm continues
to operate. For the case $\alpha<0$ but finite we shall use another
reasoning that could have also be used in the previous cases. Notice
that the characteristic frequencies of the probability amplitudes
($\sim\Omega_{0}$) are, in general, very small compared with the
time-dependent characteristic frequency of $\Omega(t)$ \textit{i.e.}
$\frac{1}{t}\int\limits _{0}^{t}\omega(t)dt$, then the stationary
phase method can be used to integrate approximately the differential
equations Eqs. (\ref{eqdiferencial1}, \ref{eqdiferencial2}) for
$\alpha\neq0$. We have used this method in the case $\alpha<0$ and
$\alpha$ finite, to obtain Eqs. (\ref{ps}, \ref{pp}) with the condition
$t<t_{c}\equiv-\frac{\gamma}{\alpha}$. The time $t_{c}$ is the `close
approach time', defined as the time when the derivative of the phase
of $\Omega(t)$ vanishes (see Eq. (\ref{omega})) and at the same
time the energy levels cross each other. Then the search algorithm
operates up to this time if $\Omega_{0}\ll\frac{1}{t}\int\limits _{0}^{t}\omega(t)dt$.
Fig. \ref{ps2} was obtained using the exact results of this paper.
It shows the probability $P_{s}$ for several values of $\alpha$,
and also that the approximation made in Eqs. (\ref{ps}, \ref{pp})
remains valid for times $t<t_{c}$. For times $t\geq$ $t_{c}$ our
previous argument cannot be applied, and the periodicity of the behavior
is not clear because other frequencies are present. This figure establishes
that, as $\left\vert \alpha\right\vert \ $is decreased, the close
approach time $t_{c}$ increases; in the limit, the algorithm works
for all times. %

To close this section, let us remark that the previous results can
also be obtained without the use of the mobile base in the following
way: Let us substitute the expressions of $f(t)$ and $g(t)$ given
by the Eqs. (\ref{ftheta},\ref{gtheta}) in the equations Eq. (\ref{scho1}).
These equations, together with the normalization of the wave function
and the conditions of maximization of the amplitude of the searched
state, allow to find the time for which the probability of the search
state is maximum. The maximization conditions are $\left\vert \frac{da_{s}(t)}{dt}\right\vert =0$
y $\left\vert a_{s}(t)\right\vert =1$ and, as a result, we obtain
that they are satisfied for those values of $\tau$ such that \begin{equation}
f(\tau)=\frac{N\left(\alpha\tau+\gamma\right)}{2\sqrt{N-1}}\cos\left(2\Omega_{0}\tau+\beta\right)=0\text{.}\label{condicion}\end{equation}
 From this equation it is deduced that, if $\alpha\tau+\gamma\neq0$
then \begin{equation}
\tau=\frac{\pi N}{2\sqrt{N-1}\epsilon}l\text{,}\label{tau2}\end{equation}
 with $l=1,2,3,...$. As a result, we see that the algorithm works
in an equivalent way to the Grover algorithm for all times, if $\alpha$
is positive, and only until the close approach time if $\alpha$ is
negative.

\section{Non adiabatic algorithm II}

In this section we introduce a new idea for the searching Hamiltonian
which is based on a different choice of the functions $f(t)$ and
$g(t)$ and possesses the characteristic that one does not need to
single out a given time in order to find the searched state with a
high probability, provided the parameters of the Hamiltonian are chosen
appropriately.

Let us return to Eq. (\ref{matriz1}) and choose now $f(t)=\frac{N}{\sqrt{N-1}}$,
then we have:\begin{equation}
H(t)=\sqrt{N-1}I+\left[\begin{array}{cc}
0 & -1\\
-1 & \text{ \ \ \ }g(t)+\frac{2-N}{\sqrt{N-1}}\end{array}\right].\label{newH}\end{equation}

The first term in this expression is constant and proportional to
the identity. One can, as done in the previous section, ignore it
for the sake of solving the Schrödinger equation. Let us choose the
(2,2) matrix element in Eq. (\ref{newH}) so that it changes linearly
with time. In this way, the resulting Hamiltonian (in the fixed basis)
mimics the evolution of the functions $a_{+}(t)$ and $a_{-}(t)$
obtained in Section 2. To be more precise, we choose\begin{equation}
g(t)=\frac{N-2}{\sqrt{N-1}}+2(b-at),\label{gII}\end{equation}
 with $a$ and $b$ constants. Notice that $f(t)$ and $g(t)$ scale
with $N$ in the same way as in the previous section, therefore the
same discussion regarding the energy cost will apply. In this second
model, the gap energy function takes the simple form $\omega(t)=\sqrt{(at-b)^{2}+1}$.

With the above definitions, apart from a global phase which we will
ignore, the Hamiltonian Eq. (\ref{newH}) gives rise to the same evolution
as the matrix\begin{equation}
H'(t)=\left[\begin{array}{cc}
0 & -1\\
-1 & 2b-2at\end{array}\right].\label{Hprime}\end{equation}
 We will allow time to run from $t=0$ to arbitrarily large values
($t\rightarrow\infty)$. As we observe, the above Hamiltonian bears
a close resemblance to the usual ones introduced in adiabatic quantum
computations, in the sense that it has a time variation which is linear
in time. However, in our case we will not start from the ground state
of the Hamiltonian, and we will not intend either to force the system
to be driven to its ground state for some finite time $T$ by making
use of the adiabatic theorem.

The resulting evolution equations for $a_{s}(t)$ and $a_{p}(t)$
can be easily decoupled, leading to

\begin{equation}
\frac{d^{2}a_{s}}{dt^{2}}+2i(b-at)\frac{da_{s}}{dt}+a_{s}=0\label{eqas}\end{equation}
 This equation has to be supplemented by the initial conditions $a_{s}(0)=\sqrt{1/N}$,
$\frac{da_{s}}{dt}(t=0)=ia_{p}(0)=i\sqrt{(N-1)/N}$. With the substitution
\begin{equation}
W=a_{s}\exp[-i(\frac{1}{2}at^{2}-bt)]\label{defW}\end{equation}
 we arrive to the same equation as in Eq. (\ref{Whittaker}), but
now $z=\sqrt{2ai}(t-b/a)$ and $\eta=-\frac{i}{2a}$. The solution
to this equation can still be written in the form (\ref{whitt2}),
with coefficients $A_{1}$ and $A_{2}$ which have to be determined
from the initial conditions. After some algebra, we arrive to\begin{equation}
A_{1}=\frac{1}{\sqrt{N}}\frac{D_{\eta-1}(-z_{0})+\sqrt{N}q_{0}D_{\eta}(-z_{0})}{D_{\eta-1}(z_{0})D_{\eta}(-z_{0})+D_{\eta}(z_{0})D_{\eta-1}(-z_{0})}\label{coefs1}\end{equation}
 \begin{equation}
A_{2}=\frac{1}{\sqrt{N}}\frac{D_{\eta-1}(z_{0})-\sqrt{N}q_{0}D_{\eta}(z_{0})}{D_{\eta-1}(z_{0})D_{\eta}(-z_{0})+D_{\eta}(z_{0})D_{\eta-1}(-z_{0})},\label{coefs2}\end{equation}
 where $q_{0}=\sqrt{2a}\sqrt{\frac{N-1}{N}}\exp(i3\pi/4)$ and $z_{0}=-b\sqrt{\frac{2}{a}}\exp(i\pi/4)$.
In order to give a result for the searched probability $P_{s}(t)=|a_{s}(t)|^{2}$,
we need to particularize the values of $N$, $a$ and $b$. Using
the asymptotic form for the parabolic cylinder functions \cite{Gradshteyn},
one can obtain the following result for the limit $\lim_{t\rightarrow\infty}P_{s}(t)$:
\begin{equation}
p(a,b)=\lim_{t\rightarrow\infty}P_{s}(t)=|A_{1}e^{k\pi/4}+A_{2}e^{-3\pi k/4}|^{2},\label{pab}\end{equation}
 with $k=\frac{1}{2a}$.

\begin{figure}
\includegraphics[scale=0.7]{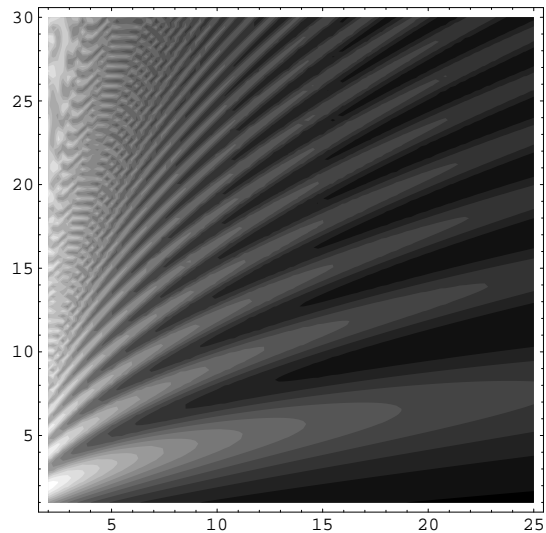}

\caption{Contour plot for the limiting probability $p(a,b)$ when $N=100$.
The horizontal axis corresponds to $a$, and the vertical axis to
$b$. Brighter regions represent a higher probability, while dark
regions indicate a probability close to zero.}

\label{fig4n100} 
\end{figure}

\begin{figure}
\includegraphics[scale=0.7]{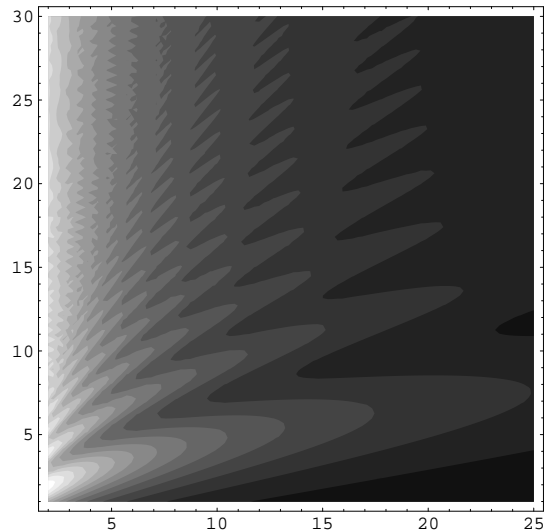}

\caption{Same as Fig. \ref{fig4n100}, for the asymptotic $N\rightarrow\infty$
limit.}

\label{fig4ninf} 
\end{figure}

Fig. \ref{fig4n100} shows a contour plot of the limiting values of
the probability $p(a,b)$ with $N=100$. As readily seen, it reveals
a complicated pattern with bands of high probability and low-probability
valleys. These patterns depend quite weakly on $N$ and, in fact,
it is possible to obtain the limit $N\rightarrow\infty$ in Eqs. (\ref{coefs2}).
Fig. \ref{fig4ninf} corresponds to this limit. As can be seen, the
changes are moderate, showing that the asymptotic probability saturates
for large values of $N$.

\begin{figure}
\includegraphics[scale=0.35]{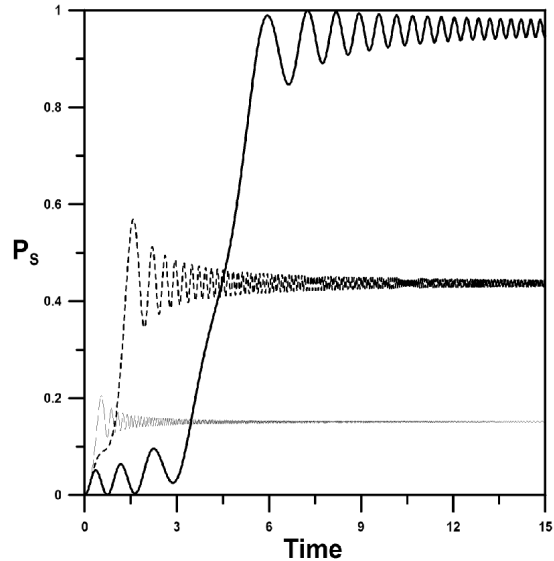}

\caption{Evolution of the probability for the searched state $P_{s}$ as a
function of time, for a fixed $b=4.5$ and three values of the parameter
$a$: 1 (solid line), 5 (thick dashes) and 20 (thin dashes).}

\label{fig5} 
\end{figure}

In order to explicitly show the differences between our proposal and
the adiabatic treatment, we have calculated the probability $P_{s}(t)$
for different parameters. The results are shown in Fig. \ref{fig5}.
Given the structure seen in Fig. \ref{fig4ninf}, we choose a fixed
$b=4.5$, and plot the probability for three values of $a$: 1, 5
and 20. The value of $N$ we used is $=10^{6}$ which, according to
the previous discussion, is equivalent to taking $N\rightarrow\infty$.

In this figure, it is apparent that a transition occurs at the time
$t_{c}=b/a$, corresponding to the minimum distance between the two
eigenvalues of the Hamiltonian. The asymptotic behavior with time
is clear from this figure, although the final probability strongly
depends on the particular choice of $a$ and $b$, as seen before.
It is important to note that the corresponding Grover time would be
of the order $\sqrt{N}\sim10^{3}$, which represents a much larger
time scale than the one showed in this figure.

Therefore, our proposal consists on implementing the Hamiltonian \ref{matriz1}
but with the functions $f$ and $g$ as defined above. By an appropriate
choice of the parameters $a$ and $b$ appearing in $g(t)$, one can
make the probability of the searched state to reach a value close
to unity within a scale of time which is much shorter (for large $N$)
than the corresponding Grover's time. At first sight, this appears
to be in contradiction with the well-known result that Grover's algorithm
is optimal for quantum searching \cite{Boyer}. However, it has been
discussed that adiabatic search can be done at a shorter time (even
with a time scale which is independent of $N$) at the expense of
increasing the energy resources \cite{Das,Wei}. This, in fact, seems
to be the case within our proposal, if one remembers that, in order
to obtain the Hamiltonian \ref{newH} using the resources defined
in \ref{matriz1}, the functions $f$ and $g$ will scale as $\sqrt{N}$,
thus increasing the energy resources as $N$ grows.

\section{Conclusions}

We have developed two new continuous time quantum search algorithms
using a time-dependent Hamiltonians in a non adiabatic regime. Our
approach differs from to the usual (adiabatic) approach, when one
starts from the initial ground state and tries to evolve slowly, making
use of the adiabatic theorem to stay close to the instantaneous ground
state. For the first case, the key of the algorithm is that the derivative
of the amplitudes $a_{-}(t)$ and $a_{+}(t)$ have a fast time variation
with a vanishing mean value over the characteristic time $\tau$ \textit{i.e.}$\frac{1}{\tau}\int\limits _{0}^{\tau}\frac{da_{_{\pm}}(t)}{dt}dt\sim0$,
then starting from the ground state, in the mobile basis $\left\{ |E_{+},t\rangle,|E_{-},t\rangle\right\} $,
the system remains near this ground state for all times. This algorithm
behaves like the Grover algorithm for non negative values of the parameter
$\alpha$, independently of its particular value, for a large $N$
and $\Omega_{0}\lll1$. The optimal search time is proportional to
$\sqrt{N}$, and the probability to find the searched state oscillates
periodically. For $\alpha<0$ the algorithm does not work for $t\geq t_{c}$,
with $t_{c}$ the close approach time.

The second algorithm makes use of similar resources to build up a
Hamiltonian that changes linearly with time. In our proposal, the
initial and final states do not correspond to the ground states of
the Hamiltonian, and the system is allowed to evolve up to arbitrarily
large times, showing a convergence towards a final state after a finite
transition time. When the parameters are chosen appropriately, the
asymptotic state can overlap with the searched state with high probability,
and one does not need to pick up a special value of time to perform
the measurement in order to obtain the desired result.

Both algorithms can be used to perform a search within a time which
can be made shorter than the standard Grover's time, at the expense
of using also larger than standard energetic resources.

These results open the possibility for the design of new quantum algorithms
that perform a search on an unstructured database (and possibly other
algorithmic tasks) alternatively to the existing algorithms.

\textbf{Acknowledgments}

We acknowledge the comments made by V. Micenmacher and the support
from PEDECIBA and PDT S/C/OP/28/84. This work has also been supported
by the Spanish Ministerio de Educación y Ciencia through Projects
AYA2004-08067-C01 and FPA2005-00711.

\end{document}